\documentclass[12pt]{article}

\usepackage[dvips]{graphicx}

\begin{document}

\newcommand{\el}{\left}
\newcommand{\er}{\right}
\newcommand{\p}{\prime}
\newcommand{\ka}{\kappa}
\newcommand{\rr}{\rho}
\newcommand{\ti}{\tilde}
\newcommand{\veps}{\varepsilon}
\newcommand{\dis}{\displaystyle}
\newcommand{\scr}{\scriptsize}

\begin{center}
{\bfseries CALCULATIONS OF THE K$^{+}$-NUCLEUS MICROSCOPIC OPTICAL
POTENTIAL AND OF THE CORRESPONDING DIFFERENTIAL ELASTIC CROSS
SECTIONS}

\vskip 5mm

V.K.LUKYANOV$^{1 \dag}$, E.V.ZEMLYANAYA$^{1}$, K.V.LUKYANOV$^{1}$,
and K.M.HANNA$^{2}$

\vskip 5mm

{\small (1) {\it Joint Institute for Nuclear Research, Dubna,
Moscow oblast, 141980 Russia }
\\
(2) {\it Math. and Theor. Phys. Dep., NRC, Atomic Energy
Authority, Cairo, Egypt }
\\
$\dag$ {\it E-mail: lukyanov@theor.jinr.ru }}
\end{center}

\vskip 5mm
\begin{center}
\begin{minipage}{150mm}
\centerline{\bf Abstract} {Calculations are made of the $K^{+}+
^{12}$C, $^{40}$Ca differential elastic scattering cross sections at
the beam momenta 0.635, 0.715, and 0.8 GeV/c. To this end the
microscopic optical potential derived in the high-energy
approximation was used where existing data on the kaon-nucleon
amplitude and on the point-like density distributions of
target-nuclei  were utilized. The effect of different methods of
relativization  was studied and shown to play an important role. A
good agreement with the experimental data on differential elastic
cross sections was obtained.}
\end{minipage}
\end{center}

\vskip 10mm

\section {Introduction and basic formulas}

It is known that the $K^+$ meson scattering can be used as a weak
hadronic interacting probe for investigating the neutron density
distributions in nuclei, and therefore the experimental data on
elastic differential cross sections at intermediate energies
\cite{Marlow},\cite{Chrien} are under the permanent attention of
theoreticians for years. For example, in \cite{Abgrall}, the
high-energy multiple scattering theory \cite{Glauber} was applied to
explain the data of the 0.8 GeV/c kaon scattering on $^{12}$C and
$^{40}$Ca, and conclusions were made that accounting for a number of
correlations in the theory did not affect on the ordinary
calculations to improve  an agreement with experimental data. Later
on, in \cite{Ebrahim1},
 calculations were made using the potential suggested in \cite{JS} as a local
version of the Kisslinger nonlocal potential \cite{Kiss}, early
constructed for the pion-nucleus scattering. The agreement with the
data was obtained if one increases  the $K^+N$ S11 phase shift by
about 10-15$\%$ . Also, the detailed analysis of the nuclear $K^+$
scattering was done in \cite{Ebrahim2} in the framework of a
phenomenological Woods--Saxon potential, depended on six free
parameters. Four sets of parameters was exhibited that minimize
$\chi^2$ for each kaon momentum 0.635, 0.715 and 0.8 GeV/c in
scattering on the $^{12}$C and $^{40}$Ca nuclei.

 In the present
work, we study the possibility to apply the microscopic optical
potential, derived in \cite{LZL} basing on the the so-called optical
limit of the high-energy multiple scattering theory:
\begin{equation}\label{eq1}
U^H=V^H+iW^H=-{{\hbar}c\,\beta\over(2\pi)^2}\,\sum_{\nu=p,n}\,{\bar\sigma}^{\nu}_{K}
({\bar\alpha}_{K}^{\nu}+i)\, \int\limits_0^\infty
dq~q^2j_0(qr)\rr_{\nu}(q)f_{K}^{\nu}(q).
\end{equation}
Here $\rr_n=\rr/2$ and $\rr_p=\rr/2$ are form factors of the bare
neutron and proton densities of a target nucleus, and $\rr$, the
corresponding form factor of the point-like charge density
distributions of nuclei \cite{LZS} obtained from the experimental
electron scattering nuclear  form factors in \cite{BKLP}. Also,
the total kaon-nucleon cross section ${\bar\sigma}^{\nu}_K$
($\nu=p,n$) and the ratio of the real to imaginary part
${\bar\alpha}^{\nu}_K$ of the $K^+N$ scattering amplitude at the
forward direction together with its form factor
$f_K^\nu=\exp(-\beta_\nu q^2/2)$ are known from \cite{Ya}.
Therefore, the potential (\ref{eq2}) has no free parameters. In
the considered processes one should use relativistic kinematics,
because of  the kinetic energies of kaons are comparable with
their rest masses. So, $\beta= k^{lab}/E$ is the laboratory
velocity and $E=\sqrt{{k^{lab}}^2+m_1^2}$ the total energy of the
incident particle $m_1$ with the momentum $k^{lab}$.

Calculations are made of the relativistic wave equation obtained
from the Klein-Gordon equation where terms quadratic in the
potential have been neglected:
\begin{equation}\label{eq2}
(\Delta\,+\, k^2)\psi({\bf r})
\,=\,2\mu\gamma^{(r)}(U\,+\,U_C)\psi({\bf r}).
\end{equation}
This form coincides with the Schr{\"o}dinger equation, but here $k$
is the relativistic momentum in the center-of-mass system. As to the
right hand side of equation, the value $\mu\gamma^{(r)}$ can be
regarded as the relativistic reduced energy (or mass) in the c.m.
system, while $\mu=m_1m_2/(m_1+m_2)$. So, we can solve this equation
as an non-relativistic one with the effective potential $U_{eff}=
\gamma^{(r)}(U^H+U_C)$, where factor $\gamma^{(r)}$ arrives because
of relativization. This latter becomes equal to 1 in the
non-relativistic limit. In fact, we apply the standard programm
DWUCK4 \cite{Kunz} for the Schr{\"o}dinger equation where one should
input the potential $U_{eff}$ and effective kinetic energy
$T^{lab}_{non}= [(m_1+m_2)/m_2]T^{cm}_{non}$ with
$T^{(cm)}_{non}=k^2/2\mu$ defined by the relativistic momentum $k$.
In our case we implement the microscopic optical potential $U^H(r)$
(\ref{eq1}) and the Coulomb potential $U_C(r)$ for an uniformly
charged sphere.

The relativistic momentum of a kaon in c.m. system is as follows
\begin{equation}\label{eq3}
k={m_2k^{lab}\over\sqrt{(m_1+m_2)^2+2m_2T^{lab}}},
\end{equation}\\
where $T^{lab}=E-m_1$.

 For different methods of relativization,
factors $\gamma^{(r)}$ are distinct from each other. In the case
when the value $\mu\gamma^{(r)}$ is considered as the reduced energy
(see, e.g. \cite{Satchler}), relativization factor has the form
\begin{equation}\label{eq4}
\gamma^{(r1)}=
\gamma_1^{\ast}\cdot{m_1+m_2\over \gamma^\ast_1 m_1+m_2},
\end{equation}
where the Lorenz-factor in c.m. system
\begin{equation}\label{eq5}
\gamma^\ast_1=
{\gamma_1m_2+m_1\over\sqrt{2\gamma_1m_1m_2+m_1^2+m_2^2}}, \qquad
\gamma_1\,=\,{E\over m_1}.
\end{equation}

In \cite{Ing}, the use is made of
\begin{equation}\label{eq6}
\gamma^{(r2)}={k^2\over{(W-m_2)^2-m_1^2}}{{W-m_2}\over\mu},
\end{equation}
where $W=\sqrt{k^2+m_1^2}+\sqrt{k^2+m_2^2}$, the total energy in
c.m. system. Also, the form of
\begin{equation}\label{eq7}
\gamma^{(r3)}\,=\,
 {k\over \beta}\,{1\over \mu}
\end{equation}
was elaborated in \cite{FIM} and it was argued that, with this
factor, the wave equation (\ref{eq2}) reproduces the relativistic
form of the Rutherford cross section for the Coulomb potential for
a point charge. Also, we apply the factor corresponding to the
relativistic equation which has long been derived in \cite{GW},
\begin{equation}\label{eq8}
\gamma^{(r4)}={W-m_2\over W}\,{m_2\over \mu}.
\end{equation}

Below we utilize the set of gamma factors (\ref{eq4}),
(\ref{eq6})-(\ref{eq8}) to calculate elastic differential cross
sections. So, we can distinguish how do different methods of
relativization affect differential cross sections of the 0.635,
0.715 and 0.8 GeV/c kaons scattered on the $^{12}C$,\,$^{40}Ca$
nuclei (see $\gamma^{(r)}$ in Table \ref{tab1}).

\begin{table}[h]
\caption{The relativization $\gamma^{(r)}$ factors in kaon-nucleus scattering
\label{tab1}} \begin{center}
\begin{tabular}{c|c|c|c|c|c}
\hline
      reaction   & $k^{lab}$, MeV & $\gamma^{(r1)}$ & $\gamma^{(r2)}$& $\gamma^{(r3)}$& $\gamma^{(r4)}$ \\
    \hline
$K^+ + ^{12}$C & 800 & 1.7421 & 1.7883 & 1.8381 & 1.7864 \\
               & 715 & 1.6303 & 1.6679 & 1.7082 & 1.6664 \\
               & 635 & 1.5269 & 1.5571 & 1.5893 & 1.5560 \\ \hline
$K^+ + ^{40}$Ca& 800 & 1.8496 & 1.8658 & 1.8824 & 1.8656 \\ \hline
\end{tabular}
\end{center}
\end{table}

\section{Results and conclusion}

\begin{figure}[h]
 \centerline{
 \includegraphics[width=0.5\linewidth]{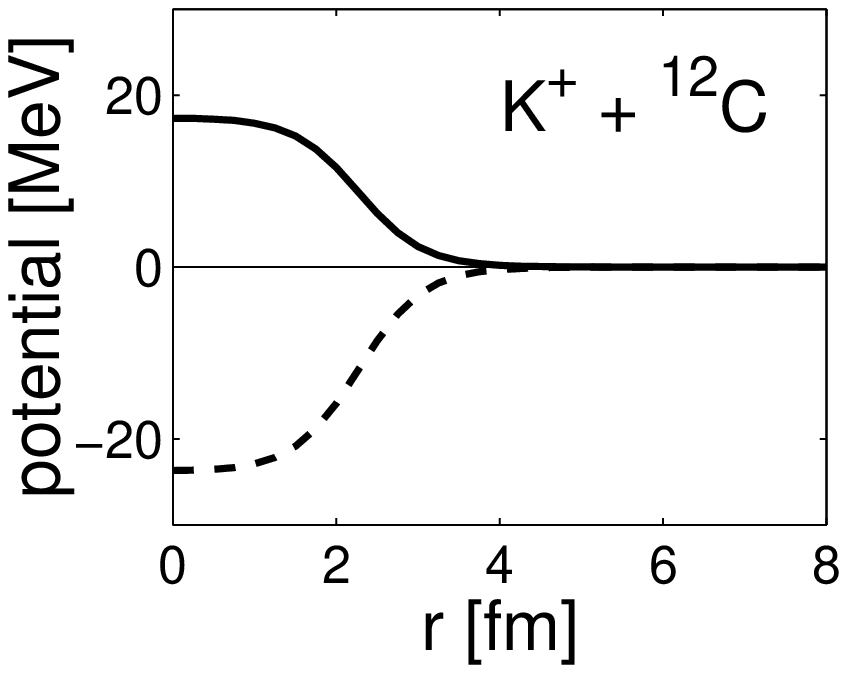}
 \includegraphics[width=0.5\linewidth]{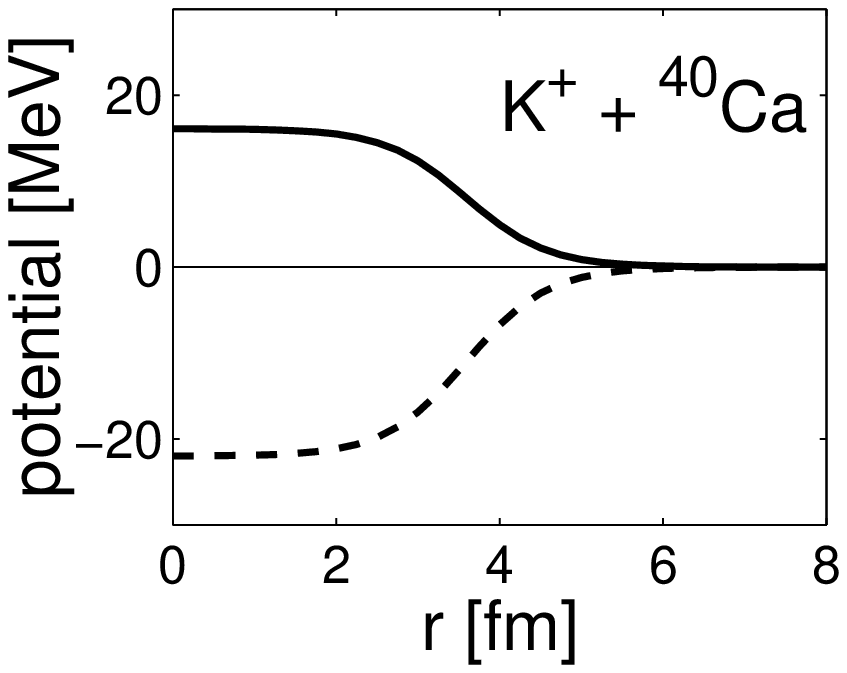}}
\caption{The real (solid lines) and imaginary (dashed lines) parts
of microscopic optical potentials (\ref{eq1}) for kaon-nucleus
scattering .\label{fig1}}
\end{figure}

First, in Fig. \ref{fig1} we exhibit as an example the behavior of
the real $V^H$ and imaginary $W^H$ microscopic potentials calculated
with a help of (\ref{eq1}) for the case of momentum 0.8 GeV/c. It is
seen the different signs of potentials, repulsive one for the real
part and attractive for imaginary one. Multiplying these microscopic
potentials by $\gamma^{(r)}$, the form of effective potentials
$U_{eff}=\gamma^{(r)}(U^H+U_C)$ are not changed while their
strengths are increased by a factor about 2.

\begin{figure}[t]
 \centerline{
 \includegraphics[width=0.5\linewidth]{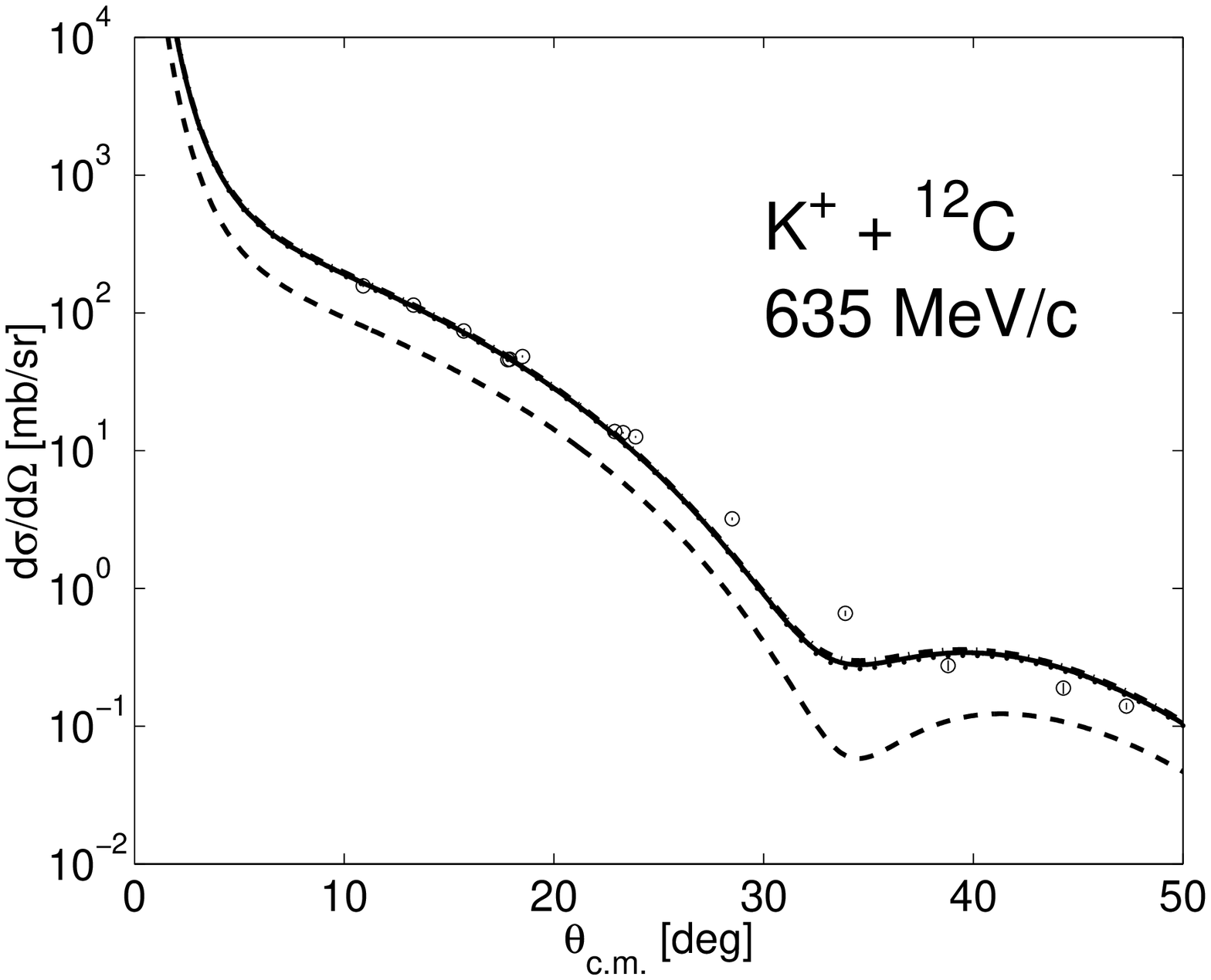}
 \includegraphics[width=0.5\linewidth]{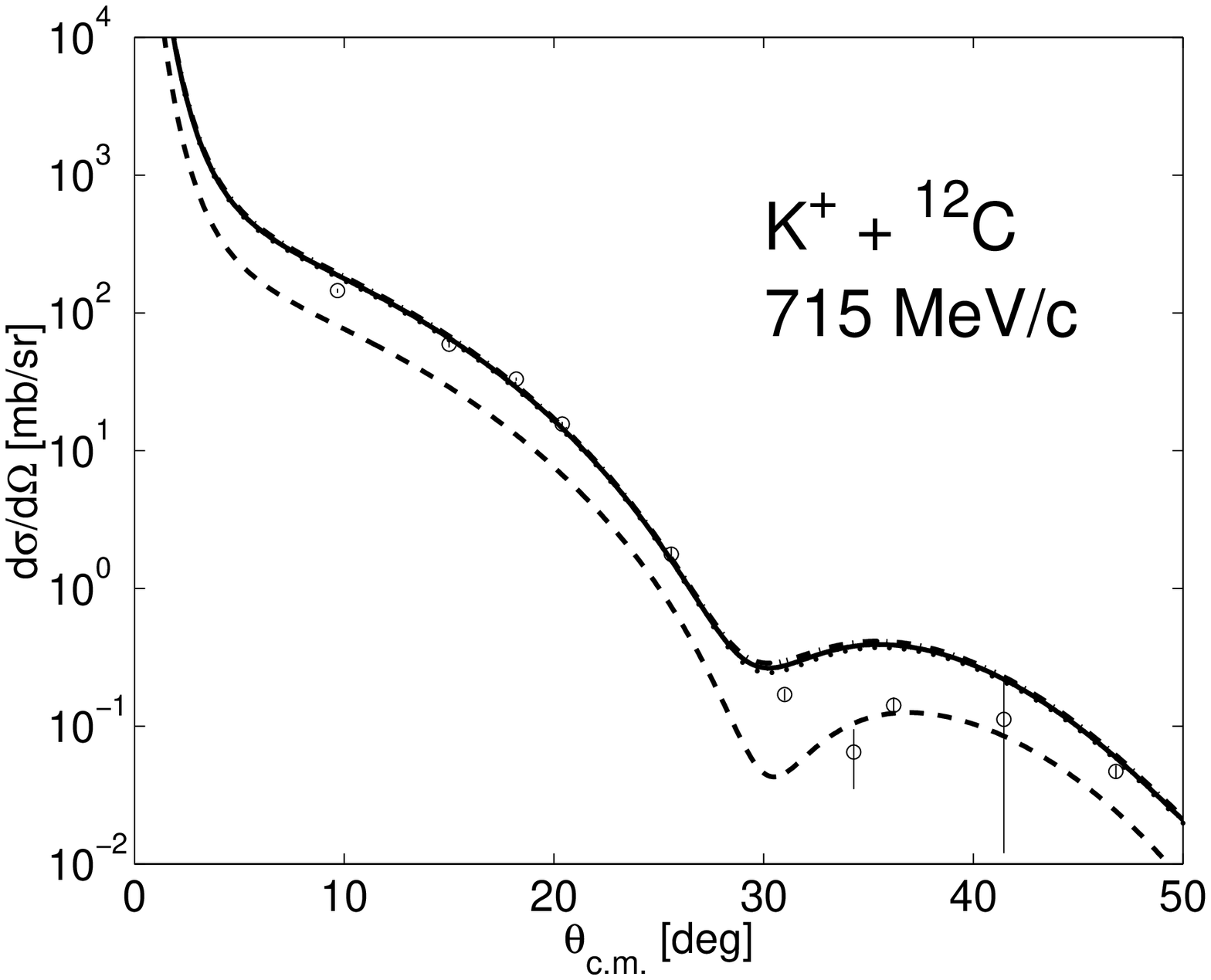}}
\centerline{
 \includegraphics[width=0.5\linewidth]{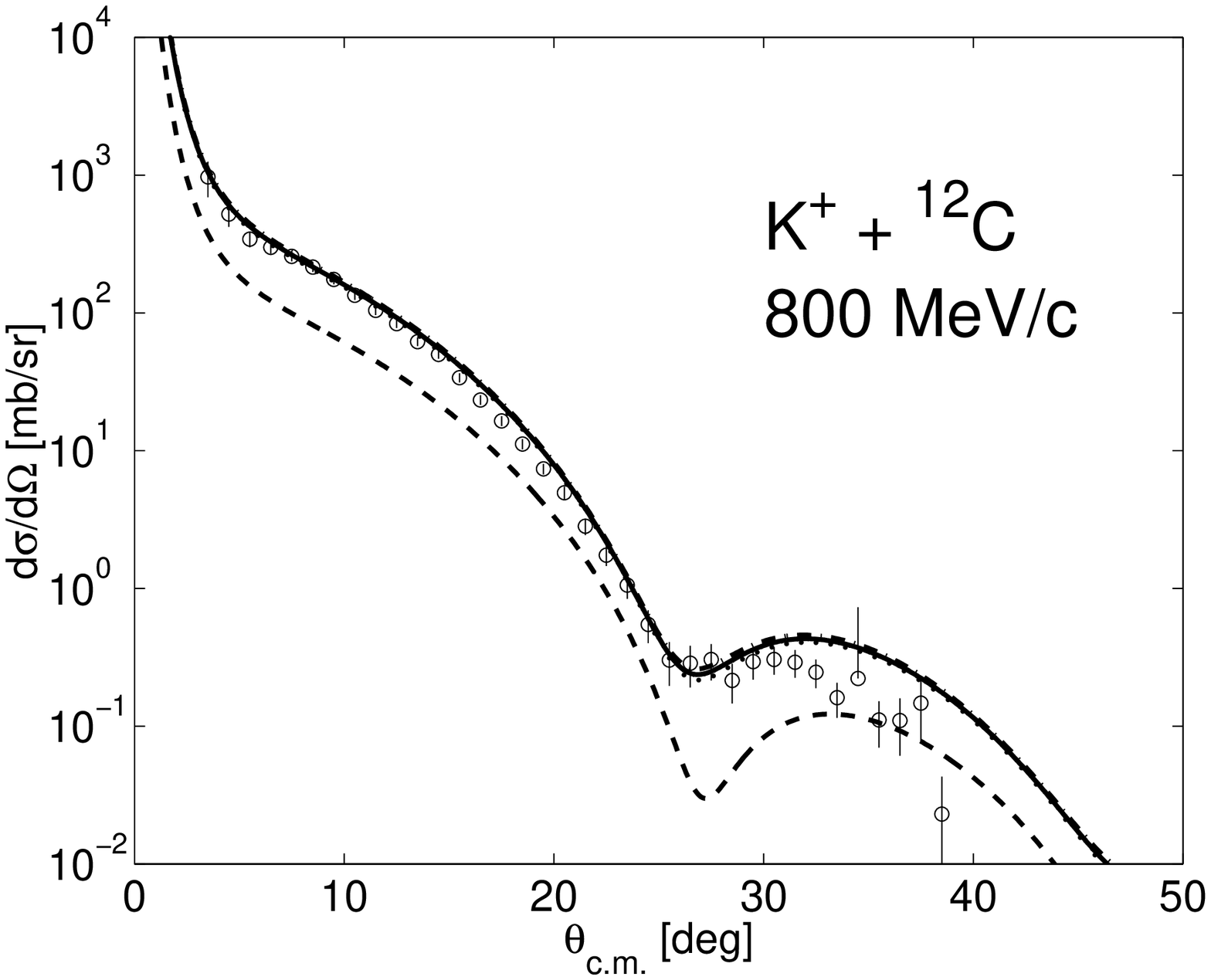}}
\caption{Differential cross sections of the  elastic scattering
$K^+\,+\,^{12}$C at different energies. Long-dashed lines correspond
to $\gamma^{(r)}=1$; short-dashed lines -- $\gamma^{(r1)}$; solid
lines -- $\gamma^{(r2)}$; dash-pointed -- $\gamma^{(r3)}$; pointed
-- $\gamma^{(r4)}$. Experimental points from \cite{Marlow,Chrien}.
\label{fig2}}
\end{figure}

\begin{figure}[h]
 \centerline{
 \includegraphics[width=0.5\linewidth]{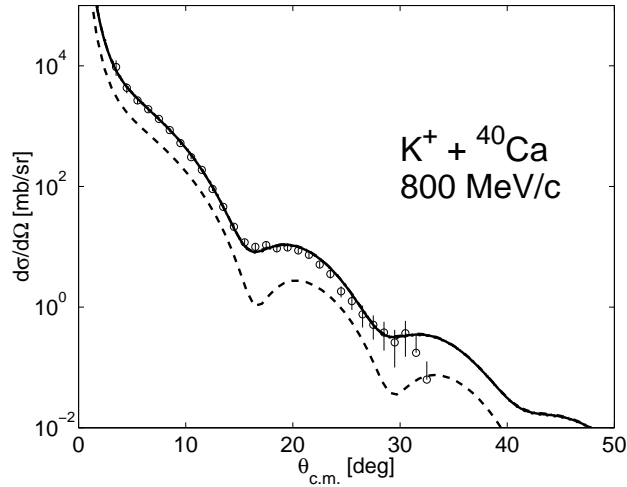}}
\caption{The same as in Fig. \ref{fig2}, but for the target nucleus
$^{40}$Ca .\label{fig3}}
\end{figure}

Figures \ref{fig2} and \ref{fig3} demonstrate differential cross
sections calculated for elastic scattering of $K^+$ on the $^{12}$C
and $^{40}$Ca nuclei. One can see that cross sections coincide very
close for each other in the considered models of relativization, and
agree  with the corresponding experimental data. The exception is
for $\gamma^{(r)}=1$ which can be called the non-relativistic case.
In fact,this is not too right to name it so, because of we have to
input the relativistic momentum $k$ (\ref{eq3}) into the equation
(\ref{eq2}). Nevertheless, one can conclude that the relativization
effects are very important when considering the kaon-nuclear
scattering at intermediate energies.


 The last item is
connected with predictions of total reaction cross sections. Here we
compare our calculations with the data presented in \cite{Friedman}
where $\sigma_R$ were estimated with a help of optical model
analysis of the measured attenuation cross sections of the $K^+$
interactions with $^{12}$C and $^{40}$Ca at momenta from 0.488 to
0.714 GeV/c. The data on $\sigma_R$ at 0.635 and 0.715 GeV/c for a
kaon interaction with $^{12}$C were estimated in \cite{Friedman}
 as about 140 and 150 $mb/sr$, while our calculations
give the corresponding values in limits of 121\,-\,125 and
126\,-\,128 $mb/sr$. These latter magnitudes are in about
10\,-\,15$\%$ less than the measured values. It is interesting to
mention that in calculations \cite{Ebrahim1} where the local
potential (equivalent to the nonlocal Kisslinger potential) was
used, the data of \cite{Friedman} were explained in a good condition
if one increases the S11 phase shift of the $K^+N$ scattering  by
about 10\,-\,15$\%$, too. So, the problem retained to get the common
experimental measurements and theoretical description of the
$K^+$-nucleus differential elastic scattering and total reaction
cross sections.

\end{document}